# Significant impact of $Al_{1-x}Ga_xN$ interlayer on GaN/AlN thermal boundary conductance

Khalid Zobaid Adnan[1], Hao Zhou[1], and Tianli Feng[1*]

[1]Department of Mechanical Engineering, University of Utah, Salt Lake City, Utah 84112, USA

**Corresponding Authors**

*Email: tianli.feng@utah.edu

**Abstract**

AlN/GaN heterostructures are central to high-power and high-frequency electronics, including RF devices, power converters, and AI accelerators. An intermediate $Al_{1-x}Ga_xN$ (AlGaN) layer is often present, either unintentionally during growth or intentionally to induce a 2D electron gas, yet its impact on the interfacial thermal boundary conductance (TBC) remains unknown due to the lack of reliable measurement or modeling methods. Here, we report a first principles-based evaluation of the TBCs of AlN/AlGaN, AlGaN/GaN, and AlN/AlGaN/GaN interfaces over the full alloy range ($0 \leq x \leq 1$). This is realized by the development of accurate deep learning interatomic potentials based on first-principles simulations. Contrary to other material systems where mixed interlayers enhance thermal coupling, we find that an AlGaN interlayer markedly degrades TBC between GaN and AlN, explaining the observation in experiments. Finally, we show that if the Al composition is sigmoidally transitioned from 0 to 1 across the AlN/GaN interface, it can remarkably increase the TBC, compared to an abrupt or a linear transition. This work is expected to shed light on an accurate thermal analysis and electro-thermal co-design of future AlGaN-based devices.

Modern high-power and high-frequency electronics[1,2], optoelectronics[3–5], and wireless communication systems[6] increasingly rely on GaN and AlN-based[7–12] high electron mobility transistors (HEMTs), owing to the wide bandgap, high electron mobility[13,14], high breakdown voltage[15,16] and excellent thermal and chemical stability. In contemporary and emerging HEMTs designs, the AlN/GaN, AlN/Al$_{1-x}$Ga$_x$N, Al$_{1-x}$Ga$_x$N/GaN, and AlN/Al$_{1-x}$Ga$_x$N/GaN heterostructures are commonly seen.[17–26] As device dimensions shrink and power density rises, the thermal boundary resistance (TBR) at these interfaces becomes a critical limiter of near-channel heat removal.[27,28] A clear understanding of interfacial thermal transport is therefore essential to push the performance limits of these devices.

The thermal management of GaN HEMTs has been widely studied in the literature.[29,30] However, most efforts focus on device-level modeling and offer limited insight into the underneath atomic-scale transport mechanisms. In particular, the thermal boundary conductance (TBC) of heterostructures involving AlGaN remains poorly understood. Additionally, simplified thermal resistance models, which treat interfaces as independent resistors and utilize the bulk thermal conductivities of each layer, are commonly employed, yet their validity breaks down at the nanoscale. Recent work by Adnan and Feng has shown that surrounding structures can significantly alter the TBC of the original interface due to ballistic effects, highlighting the need for atomistic-level studies to address those limitations.[31]

Non-equilibrium molecular dynamics (NEMD)[31–34] is an effective tool for probing interfacial thermal transport, as it naturally incorporates realistic atomic configurations and captures both elastic and inelastic phonon interactions.[35–44] However, a key limitation lies in the reliability of empirical interatomic potentials such as Tersoff or Stillinger-Weber, which often fail to capture the complex bonding environments at the interface and in graded or alloyed systems, leading to large discrepancies between predicted and measured TBC values for GaN/AlN interfaces.[45–51] The recent evolving machine learning interatomic potentials (MLIPs), which retain the efficiency of classical potentials while achieving near-first-principles accuracy, could be able to resolve this limitation.[52–55]

In this study, we investigate the TBC of AlGaN-based heterostructures using NEMD combined with a deep learning interatomic potential trained on density functional theory (DFT) data. We report the theoretical TBC values of AlN/Al$_{1-x}$Ga$_x$N, Al$_{1-x}$Ga$_x$N/GaN, and AlN/Al$_{1-x}$Ga$_x$N/GaN interfaces, and show that TBC strongly depends on the Ga composition, $x$. Finally, we find that the specific grading profile of Al$_{1-x}$Ga$_x$N plays a crucial role in determining interfacial TBC, pointing to new opportunities for optimizing thermal design in GaN-based devices. This work is expected to shed light on practical electro-thermal management engineering of AlGaN-based devices.

To investigate interfacial heat transfer in AlGaN-based heterostructures, we first developed a deep neural network interatomic potential trained on first-principles data. The training dataset was generated using *ab initio* molecular dynamics (AIMD) simulations of bulk AlN and GaN (288 atoms each) and AlN/Al$_{1-x}$Ga$_x$N and Al$_{1-x}$Ga$_x$N/GaN heterostructures (384 atoms) with alloy compositions ranging from $x = 0$ to 1. Before we do any calculations, we iteratively relax the heterostructures to minimize the stress by adjusting the simulation box size to minimize the stress. AIMD calculations are first performed in the NVT ensemble at 300K, which generates a total of 18000 atomic configurations with random atom displacements (0.05 Å), including bulk GaN, AlN, and the GaN/Al$_{1-x}$Ga$_x$N heterostructure. Subsequently, accurate energy, virial, and force calculations are performed for each configuration using the DFT implemented in the Vienna Ab initio Simulation Package (VASP)[56,57] with the projected augmented wave method (PAW)[58] with Perdew-Burke-Ernzerhof (PBE) exchange-correlation functional.[59] The plane-wave energy cutoff is set to be 500 eV while the convergence threshold for electronic energy is $2\times10^{-6}$ eV. A Γ-only k-mesh for electrons is used to do simulations with periodic boundary conditions applied to all directions. The accuracy of the trained potential is benchmarked against DFT calculations (Fig. 1), showing excellent agreement in both energy and force predictions.

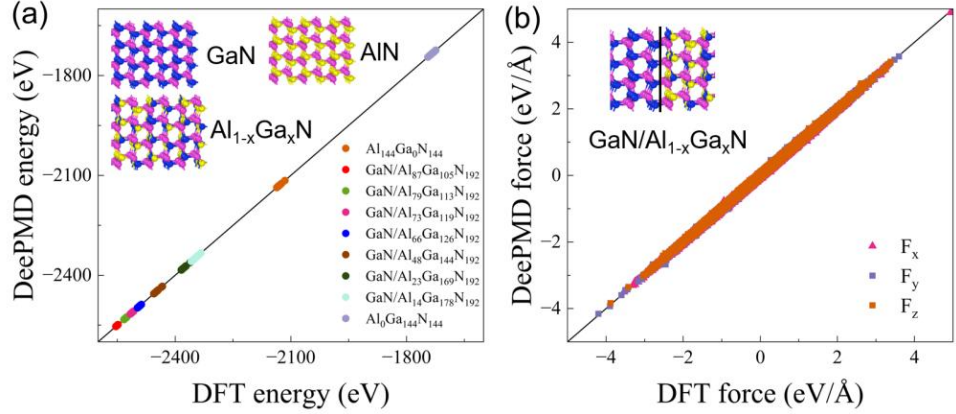

FIG 1. Comparison between DFT and DeepMD predicted (a) atomic energies and (b) forces. Both are tested on bulk GaN, bulk AlN, and heterostructures.

TBC was computed using NEMD simulations performed with the LAMMPS package.[60] We imposed shrink-wrapped boundary conditions along the heat flux direction, with fixed atoms at the system boundaries to emulate adiabatic conditions. Periodic boundary conditions were applied in the transverse directions. Each structure was equilibrated at 300 K using the NVT ensemble for 3,000,000 steps with a timestep of 0.5 fs.

The TBC values in Fig. 2 provide the first comprehensive dataset for AlN/Al$_{1-x}$Ga$_x$N or Al$_{1-x}$Ga$_x$N/GaN interfaces, obtained in this work. These values should serve as critical guidance for future thermal simulations and experiments. Experimentally, direct measurement has not yet been reported because the associated interface resistance is too small relative to other thermal resistances in typical metrology stacks, pushing current instrument sensitivity limits. On the theoretical side, first-principles calculations at these scales remain computationally prohibitive. Prior work by Polanco and Lindsay discussed possible effects of interatomic mixing on GaN/AlN TBC but could not reach a definitive conclusion.[61]

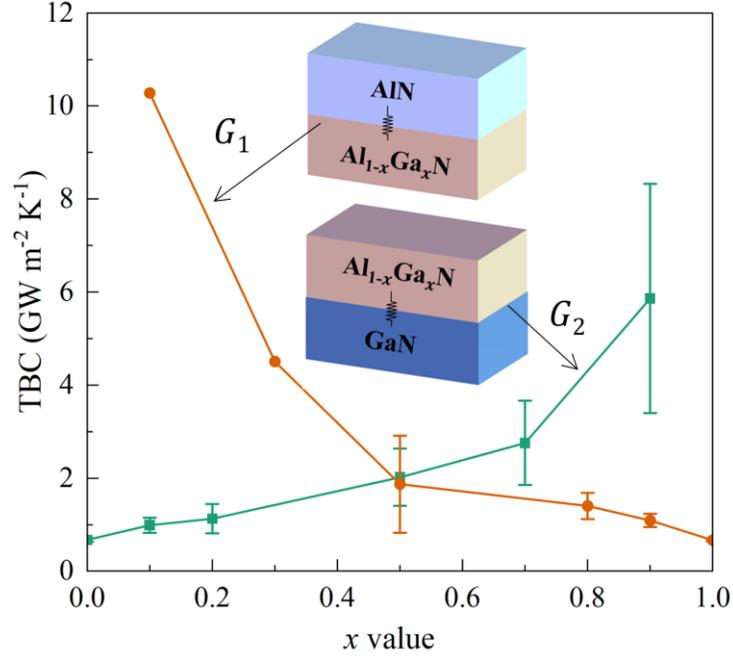

FIG 2. Room temperature quantum-corrected TBCs of standalone AlN/Al$_{1-x}$Ga$_x$N and Al$_{1-x}$Ga$_x$N/GaN interfaces as a function of composition of Ga.

When $x=1$ in AlN/Al$_{1-x}$Ga$_x$N or $x=0$ in Al$_{1-x}$Ga$_x$N/GaN, the structure reduces to an abrupt GaN/AlN interface; in both limiting cases, we obtain a TBC of ~660 MW m$^{-2}$ K$^{-1}$, consistent with our previous studies using a different machine learning potential, supporting the accuracy of our current model.[62] For comparison, GaN/AlN TBC values computed with legacy Tersoff and Stillinger-Weber potentials span from 400 to 2,000 MW m$^{-2}$ K$^{-1}$, which are not credible since these legacy potentials have never been validated against experiment or DFT for heterostructures.[62] At the opposite limit ($x=0$ in AlN/Al$_{1-x}$Ga$_x$N/GaN or $x=1$ in Al$_{1-x}$Ga$_x$N/GaN), the interface no longer exists, and the TBC becomes infinite. For intermediate compositions, $0<x<1$, the TBC is of order GW m$^{-2}$ K$^{-1}$; for example, both AlN/Al$_{0.5}$Ga$_{0.5}$N and Al$_{0.5}$Ga$_{0.5}$N/GaN yield TBCs near 2 GW m$^{-2}$ K$^{-1}$. These large TBCs, and therefore very small interfacial resistances, help explain why experimental measurement has been challenging to date. As a reference, the prior highest measured TBCs are 1.1 GW m$^{-2}$ K$^{-1}$ (for Al/MgO interface at 60 GPa) and 0.8 GW m$^{-2}$ K$^{-1}$ (for SrRuO$_3$/SrTiO$_3$ interface). [63]

It's important to note, however, that while the TBCs of AlN/AlGaN and AlGaN/GaN interfaces are high due to reduced interfacial mismatch, the AlGaN alloy layer itself introduces thermal resistance due to phonon-mass and bond disorder scattering. van Roekeghem *et al.* demonstrated using atomistic Green's function calculations that in GaN/AlN systems with graded AlGaN interlayers, phonon-alloy scattering dominated the overall thermal resistance.[64]

Interfacial atomic mixing commonly occurs at AlN/GaN interfaces, either by design or unintentionally, forming AlN/Al$_{1-x}$Ga$_x$N/GaN heterostructures. The introduction of an AlGaN layer increases or decreases the AlN/GaN TBC has remained an open question. Due to the lack of methods capable of accurately capturing the effects of bonding disorder, this issue has not been conclusively resolved. In general, interfacial mixing introduces two competing effects: (i) spectral bridging, which can enhance phonon transmission by smoothing the vibrational mismatch between adjacent crystals, and (ii) disorder, which scatters (and backscatters) phonons, thereby suppressing transmission. For Ar/heavy-Ar and Si/Ge interfaces, interfacial mixing has been shown to enhance TBC[65–71], whereas for Si/heavy-Si model systems, mixing leads to a reduction in TBC.[32]

Figure 3 presents the overall TBC of GaN/AlN interfaces with an AlGaN interlayer, denoted as $G_{\text{tot,true}}$, which includes resistances of two interfaces plus that of the AlGaN interlayer, is shown Fig. 3. Here, the subscript "true" denotes "ground truth" from direct NEMD simulations, in contrast to $G_{tot}$ predicted by the thermal circuit model. NEMD serves as "ground truth" because it naturally and implicitly captures ballistic, coherent, and coupling effects among the two interfaces and the intermediate layer, whereas the thermal circuit model treats these three resistances as independent. The $G_{\text{tot,true}}$ is evaluated from the temperature drop extrapolated to the mid-plane of the interlayer using linear fits to the adjacent AlN and GaN temperature profiles.

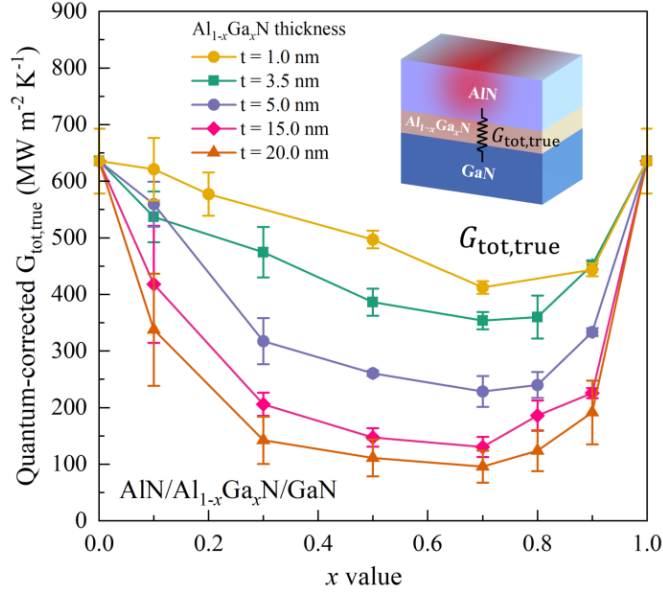

FIG. 3. Impacts of x and thickness of interlayer on (quantum-corrected) TBC of AlN/Al$_{1-x}$Ga$_x$N/GaN heterostructures obtained from machine learning molecular dynamics.

We find that introducing an AlGaN interlayer substantially reduces the overall TBC, with a stronger reduction at higher degrees of mixing and peaking near $x_{Ga}$ = 0.7. For a 1-nm interlayer, Al$_{0.8}$Ga$_{0.2}$N, Al$_{0.5}$Ga$_{0.5}$N, and Al$_{0.3}$Ga$_{0.7}$N reduce the TBC to 575, 500, and 400 MW m$^{-2}$K$^{-1}$, respectively.[72] Increasing thickness further amplifies the reduction: at $t$ = 5 nm, the same compositions yield 550, 250, and 220 MW m$^{-2}$ K$^{-1}$; at $t$ = 20 nm, the TBC falls further to 330, 110, and 90 MW m$^{-2}$ K$^{-1}$, respectively. This large decrease contrasts with prior expectations, as interfacial intermixing at lattice-matched interfaces (e.g., Ar/heavy-Ar and Si/Ge) is often found to enhance TBC[65–71]. The observed TBC degradation likely arises from disorder-induced phonon backscattering within the interlayer. Because AlN and GaN are closely lattice-matched, the benefit from spectral bridging is modest, whereas mass and force-constant disorder in Al$_{1-x}$Ga$_x$N introduces strong alloy scattering. Moreover, the interlayer itself adds an additional series resistance, as Al$_{1-x}$Ga$_x$N possesses a substantially lower thermal conductivity than either AlN or GaN.[73,74]

These findings may explain the recent experimental observations. While simulations using either MTP or DeepMD potentials predict ~600–700 MW m$^{-2}$ K$^{-1}$ for a pristine GaN/AlN interface, measurements consistently report lower values: 320 MW m$^{-2}$ K$^{-1}$ (Li *et al.* [75]) and 490 MW m$^{-2}$ K$^{-1}$ (Walwil *et al.* [76]). A likely cause is an intermixed layer at the interface. Indeed, high-resolution transmission electron microscopy in Li et al. reveals a ~3-nm interfacial region suggestive of atomic mixing, consistent with a thin disordered AlGaN layer. Our results show that even a 3 nm disordered AlGaN interlayer can depress the effective TBC to below ~340 MW m$^{-2}$ K$^{-1}$, in line with experiment. By contrast, the GaN/AlN interface studied by Walwil *et al.* appears cleaner, with less interfacial mixing, and therefore exhibits a higher TBC closer to the pristine value. Additional structural defects at the interface may also influence the measured TBC and further contribute to values below the pristine limit.

Finally, we investigate how the distribution of Ga and Al atoms in the AlN/Al$_{1-x}$Ga$_x$N/GaN structure affects the TBC of the whole structure. Figure 4 shows the quantum-corrected TBC as a function of the interlayer thickness for Al$_{1-x}$Ga$_x$N with either a linearly or sigmoidally graded Ga composition profile. In both cases, the TBC decreases with increasing interlayer thickness due to the added thermal resistance. However, this decline is much more pronounced for the linear profile. This is attributed to sharper compositional changes in the linear grading, which introduce stronger vibrational mismatch across the interlayer. In contrast, the smoother variation in the sigmoid profile better preserves phonon continuity, enabling more efficient transmission. A similar trend was reported by van Roekeghem *et al.* using the atomistic Green's function method. [64] These results suggest that the spatial distribution of alloy atoms plays a key role in governing interfacial thermal transport and can be engineered to enhance the thermal management performance of AlGaN-based high electron mobility transistors.

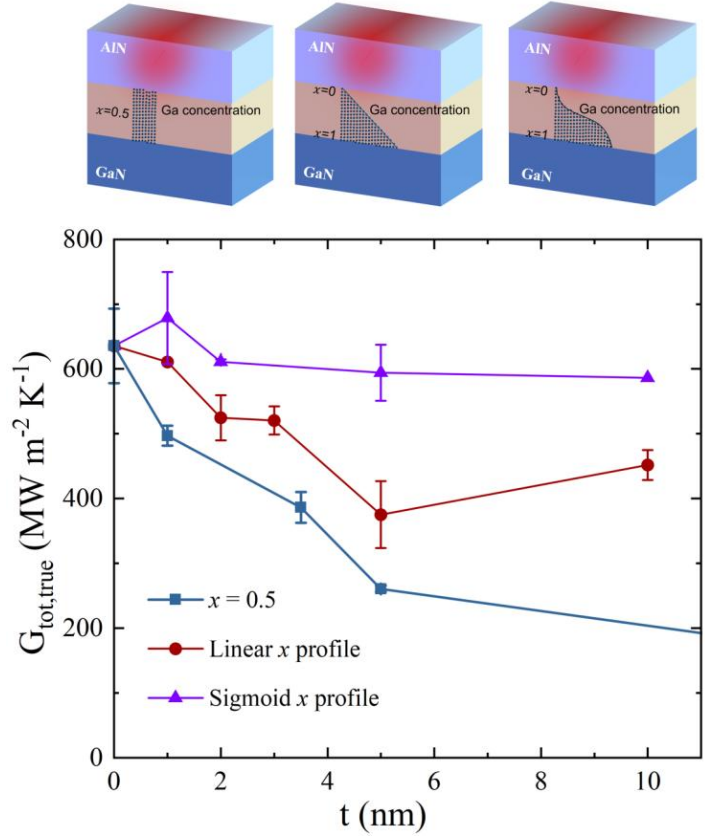

FIG. 4. Impact of graded interface profiles on TBC of AlN/Al$_{1-x}$Ga$_x$N/GaN heterostructures.

In conclusion, we have employed deep learning interatomic potential trained on first-principles data to systematically investigate interfacial thermal transport in AlGaN-based heterostructures. Our NEMD simulations revealed that while the TBC values of AlGaN/GaN and AlGaN/AlN interfaces are negligible, inserting an AlGaN interlayer between GaN and AlN significantly reduces the overall TBC, primarily due to phonon-alloy scattering that reduces the original high phonon transmission coefficients across the lattice-matched AlN/GaN interface. Finally, we showed that compositionally graded interlayers, especially with smooth sigmoid-like profiles, can help preserve TBC. These insights provide critical guidance for the thermal design of AlGaN-based electronic devices.

**Data availability**

Source data are provided with this paper. All other data that support the plots within this paper are available from the corresponding authors on reasonable request.

**Code availability**

The codes used in this study are available from the corresponding authors upon request.

**Acknowledgments**

This work is supported by National Science Foundation under Award 2337749. We thank the computational support from the Center for High Performance Computing (CHPC) at the University of Utah and Bridges-2 at Pittsburgh Supercomputing Center through allocation MCH240097 from the Advanced Cyberinfrastructure Coordination Ecosystem: Services & Support (ACCESS) program. T. Feng thanks Dr. Mahesh R Neupane at Army Research Laboratory for fruitful discussions.

**Competing interests**

The authors declare no competing interests.

**Additional information**

Correspondence and requests for materials should be addressed to T.F.